\documentclass{osa-article}
\usepackage{float}

\journal{oe}

\articletype{Research Article}

\begin{document}

\title{Phase-contrast THz-CT for non-destructive testing}

\author{Peter Fosodeder,\authormark{1,*} Simon Hubmer,\authormark{2} Alexander Ploier,\authormark{2,3} Ronny Ramlau,\authormark{2,4} Sandrine van Frank\authormark{1} and Christian Rankl\authormark{1}}

\address{
\authormark{1}Research Center for Non Destructive Testing GmbH (RECENDT), Altenbergerstraße 69, A-4040 Linz, Austria
\\
\authormark{2}Johann Radon Institute Linz, Altenbergerstra{\ss}e 69, A-4040 Linz, Austria
\\
\authormark{3}Doctoral Program Computational Mathematics, Johannes Kepler University Linz, Altenbergerstra{\ss}e 69, A-4040 Linz, Austria
\\
\authormark{4}Institute of Industrial Mathematics, Johannes Kepler University Linz, Altenbergerstra{\ss}e 69, A-4040 Linz
}

\email{\authormark{*}peter.fosodeder@recendt.at} 



\begin{abstract}
A new approach for image reconstruction in THz computed tomography (THz-CT) is presented. Based on a geometrical optics model containing the THz signal amplitude and phase, a novel algorithm for extracting an average phase from the measured THz signals is derived. Applying the algorithm results in a phase-contrast sinogram, which is further used for image reconstruction. For experimental validation, a fast THz time-domain spectrometer (THz-TDS) in transmission geometry is employed, enabling CT measurements within several minutes. Quantitative evaluation of reconstructed 3D printed plastic profiles reveals the potential of our approach in non-destructive testing of plastic profiles.
\end{abstract}

\section{Introduction}

Over the past years, THz technology has evolved into a widely recognized technology for industrial non-destructive testing (NDT), its most prominent application being the thickness control of thin coating layers \cite{Jonuscheit_THzLayerThickness, Tu_Zhong_MarineProtectiveCoatings, Su_AutomotiveCoatings}. Equally important and well-established is the application of THz spectroscopy \cite{Jin_Kim_Jeon_THzDielectricConstants_2006, Pupeza_TeraLyzer, Scheller_TeraLyzer} in chemical process control \cite{Strachan_THzTDSPharmaceutical} or even in security applications \cite{Shen_ExplosivesTHzSpectroscopy}. In this work, we concentrate on one of the less established NDT methods, namely THz computed tomography (THz-CT). The following paragraphs contain a short motivation for the use of THz-CT, a description of the state-of-the-art, current challenges and a description of our approach for extending the applicability of THz-CT.

Computed tomography (CT) is an imaging technique which uses multiple transmission measurements from different angles to produce tomographic (cross-sectional) images. Most prominently, X-radiation \cite{Garcea_Wang_Withers_XCTforComposites_2018} or ultrasound \cite{Choi_USCT, Wiskin_USCT_fullwave} is used to probe the sample. We are particularly interested in using THz radiation for CT, since it provides industrially relevant advantages over the complementary NDT methods mentioned above. THz radiation is completely harmless to humans and does not require extensive shielding, which is necessary e.g. in X-Ray based systems. In addition, THz measurements are operated contact-less, without the need for coupling media, which are required in ultrasound testing. In the past, limitations concerning the measurement speed represented a major obstacle for the use of THz systems in industrial environments. However, the current development of fast and compact measurement systems \cite{Dietz_Vieweg_ECOPS_2014} provides a suitable solution to this challenge. We propose a potential application of THz-CT for in-line quality control of plastic profiles. Applied within such a major industrial sector, the system demonstrated in this work could provide immediate feedback regarding the quality of produced plastic profiles and therefore effectively reduce the amount of produced plastic waste in case of production failure. 

Due to demands for high output power or the early limitations of measurement speed of pulsed systems, existing THz-CT approaches are often based on continuous wave (cw) source \cite{Recur_cwTHzCT_2012, Trichopoulos_5kpxTHzCT} and detector technology. However, one major drawback regarding the use of cw systems is their low operating frequency below several hundred GHz, resulting in a low diffraction limited resolution. Furthermore, conventional cw systems are typically restricted to imaging based on the transmitted intensity only. On the other hand, approaches based on pulsed technology \cite{Neu_Schmuttenmaer_THzTDS_2018, Ferguson_TRayCT, Abraham_RefractionLossTHzCT, Tripathi_THzCT_2016} provide access to the THz pulse phase and allow imaging with frequencies up to several THz, thus providing more comprehensive information about the sample.

Regarding the reconstruction of cross-sectional images, most of the existing approaches, including the ones reported in the literature (cw and pulsed) cited above, utilize methods developed for X-Ray CT. These methods are typically based on the Radon imaging model \cite{Natterer_2001} and thus do not account for the physical effects of electromagnetic wave propagation like scattering, diffraction and refraction. Approaches to reconstruct images based on more generally valid imaging models such as Snell's law of refraction \cite{Tepe_ART_THzCT_refraction} or Helmholtz's equation \cite{WaldSchuster_THzCT_Helmholtz} have been reported to improve image reconstruction. This however comes at the cost of having to introduce preliminary knowledge about the sample geometry and/or a significantly increased computational effort. 

In this paper, we present an approach for THz-CT based on a fast THz time-domain spectrometer. Based on geometrical optics, we formulate the imaging model and present two methods for calculating a sinogram from the measured THz signals. Analogously to X-Ray CT, the first method utilizes amplitude information only, while the second method exploits the fact that in THz-TDS the signal is sampled with sub-picosecond resolution, which allows us to extract a pulse phase sinogram. Cross-sectional images are reconstructed from such sinograms using different image reconstruction approaches, namely filtered backprojection, Tikhonov regularization and Landweber iteration \cite{Natterer_2001,Louarn_Verinaud_Korkiakoski_Hubin_Marchetti_2006,Engl_Hanke_Neubauer_1996}. The performance of our approach was tested by measuring 3D-printed plastic profiles and quantitatively comparing the outcome of the above procedure with the known geometry.

The goal of this article is to demonstrate the robustness, accuracy and therefore the potential applicability of our THz-CT approach to in-line NDT of extruded plastic profiles.

\section{Materials and methods}

\subsection{The THz-TDS setup}

Our setup is based on the TeraFlash smart \cite{Dietz_Vieweg_ECOPS_2014} THz-TDS system (TOPTICA, Munich). This system uses two photoconductive antennas (PCA) and two synchronized femtosecond lasers, operating at a center wavelength of 1560 nm for generating and detecting pulsed THz radiation.

\begin{figure}[h!]
    \centering
    \includegraphics[viewport=0bp 560bp 595bp 810bp,clip,width=13cm]{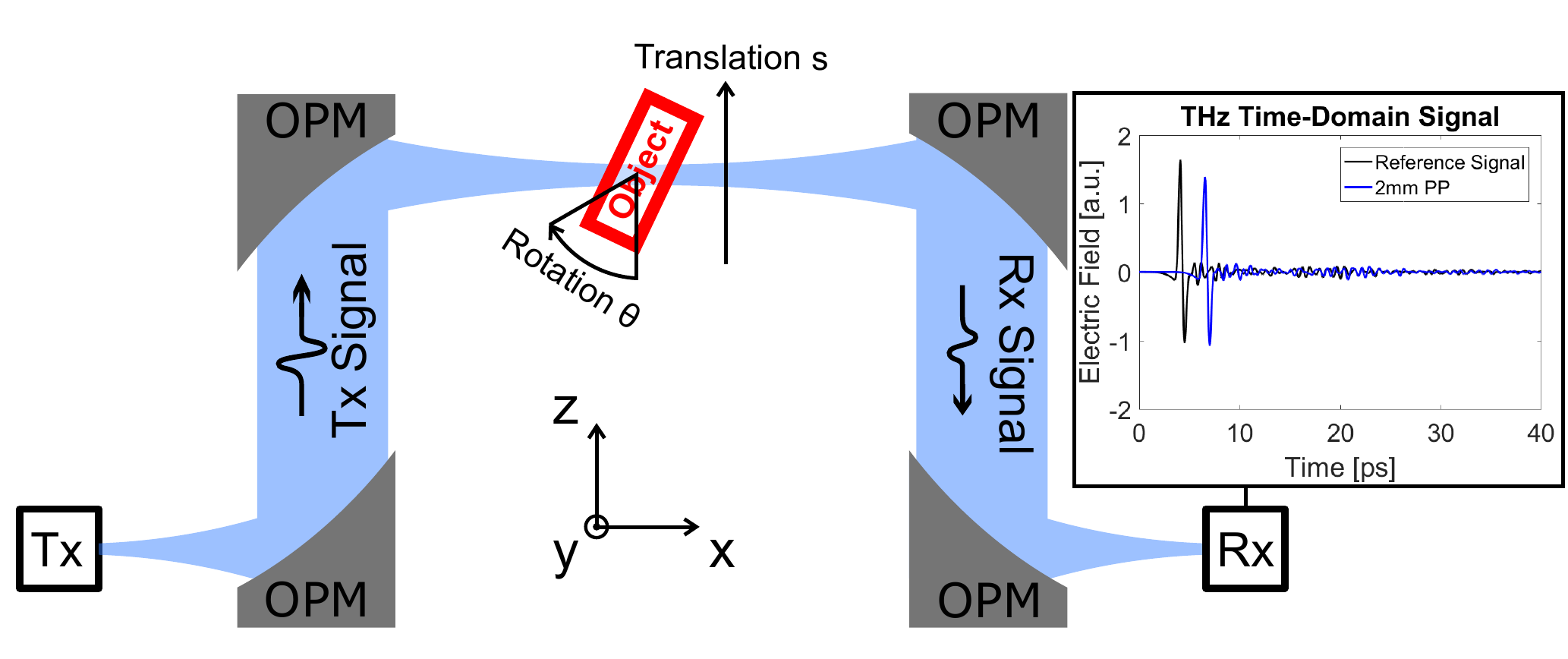}
    \caption{Schematic drawing of the measurement setup. The THz radiation is generated by a PCA (Tx). Two Off-Axis Parabolic Mirrors (OPM) are used to form a focused THz beam. After interacting with an object, the THz beam is guided to the detector PCA (Rx) by two OPMs again. Sample movement is performed by a motorized translation and rotation stage. Exemplarily, a measured reference signal through air and a measured signal through a 2 mm plate of PP is shown.}
    \label{fig_ImagingSetup}
\end{figure}

As illustrated in Fig.~\ref{fig_ImagingSetup}, the THz pulses are guided and focused by 4 off-axis parabolic mirrors (OPM), and the sample is mounted in the focal plane. Sample positioning for the CT measurement is performed by one translation and one rotation stage. For imaging in transmission it is crucial to find a reasonable trade-off between depth of field and focal spot-size, therefore the effective focal length of the focusing OPM was dimensioned with 101.6 mm. A Gaussian focal spot with a FWHM of approximately 1 mm (measured by the knife-edge technique) was achieved while still preserving enough depth of field for sample diameters of several centimeters.

Fig. \ref{fig_ImagingSetup} also illustrates two THz signals, one reference measurement in air and one measurement on a 3D printed plate of Polypropylene (PP) with a thickness of 2 mm. Typically, the bandwidth of a single shot THz reference signal exceeds 3.5 THz at a measurement rate of 1600 signals per second. The total length of one THz signal is 152.3 ps, sampled with a resolution of $\approx$ 49 fs.

\subsection{Sample preparation}

\begin{figure}[h!]
    \centering
    \includegraphics[viewport=20bp 260bp 540bp 570bp,clip,width=8cm]{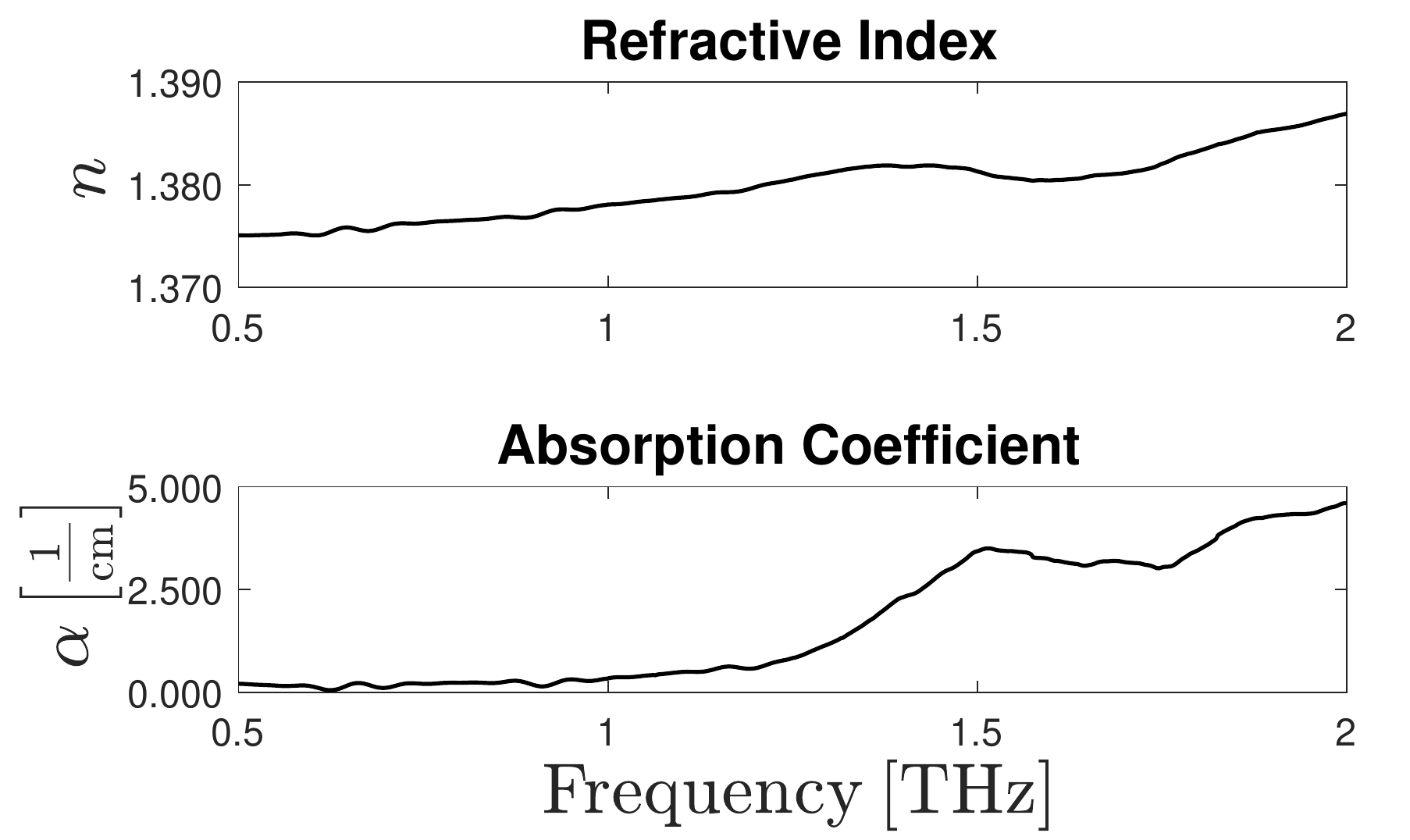}
    \caption{Material parameters of a 3D printed plate of PP (thickness = 2 mm). The absorption coefficient $\alpha$ and the refractive index $n$ were calculated from a THz-TDS measurement in transmission. No significant absorption features caused by potential structural inhomogenities were found.}
    \label{fig_SamplePreparation}
\end{figure}

The samples for this study were 3D printed from natural polypropylene (PP) at a temperature of 240 \textdegree{}C. Due to minor visible structural inhomogenities from the 3D printers layer resolution of 150 µm, a THz spectroscopy experiment on a printed plate of PP was performed. The results in Fig.  \ref{fig_SamplePreparation} were calculated, using a frequency-domain model-based parameter estimation \cite{Pupeza_TeraLyzer, Scheller_TeraLyzer}. Despite the visible surface structure of the printed samples, no significant absorption features compared to other THz spectroscopy experiments performed on PP were found \cite{Jin_Kim_Jeon_THzDielectricConstants_2006}.

\subsection{Computed tomography (CT) measurements}\label{subsec_CTMeasurements}

In conventional X-Ray CT measurements, detector arrays are used in order to minimize acquisition time. Since our THz-TDS setup employs two PCAs, effectively acting as a static point-like emitter/detector pair, we perform continuous line scans by moving the sample through the THz focus, at typical velocities of 150 mm/s. After each linescan, the sample is rotated by 1\textdegree{} until the full 360\textdegree{} scan is completed. Given the typical linescan velocity of 150 mm/s and a measurement rate of 1600 THz signals per second, a spatial discretization smaller than 0.1 mm and an angular discretization of 1\textdegree{} could be achieved in a total acquisition time of less than 5 minutes per scan.

\subsection{Data processing algorithms}\label{subsec_preprocessing}

In order to improve the signal-to-noise ratio (SNR) of the acquired THz signals, a digital FIR low-pass filter \cite{Shenoi_DigitalSignalProcessing_2005} with a passband frequency of 2.5 THz and a stopband frequency of 3.5 THz (attenuation: -80 dB) was applied to the raw data. 

Consecutively acquired THz signals were filtered by a moving average filter with a window size of five signals. This effectively corresponds to a spatial low-pass filtering of the dataset and therefore further reduces the SNR.

Due to the continuous motion of the sample, the acquired measurement positions are available on non-equidistantly distributed positions only. After calculating the sinogram values on the corresponding positions (see section~\ref{subsec_PulseEnergySinogramMapping}), the values were interpolated on an equidistant grid with a spatial discretization of 0.1 mm and an angular discretization of 1\textdegree{}.

After image reconstruction, the retrieved images were thresholded in order to remove values smaller than zero. These values typically appear due to numerical reasons and are nonphysical.

\section{The THz imaging model}

The basis for image reconstruction is an accurate and ideally compact forward model, that will be derived within this section. In its most general form, the propagation of electromagnetic radiation through an object is described by the macroscopic Maxwell equations \cite{Griffiths_Electrodynamics_2014}. In this framework, the common phenomena of wave propagation, such as refraction and diffraction, are considered. Recently, efforts have been taken in order to perform THz CT image reconstruction within this framework \cite{WaldSchuster_THzCT_Helmholtz}.

Since our work is specifically dedicated to 3D imaging of plastic profiles, some assumptions will be introduced in the following, leading to a more compact imaging model for our specific problem. The basis of our imaging model is the well-known regime of geometrical optics, where a light beam is treated as a mathematical ray, undergoing reflection, refraction and absorption by the sample. In this regime, the material-light interaction is commonly assumed to be linear and isotropic. Furthermore, the appearance of multiple reflections within a sample is neglected, since they result only in minor contributions to the measured signal.

Since the samples in this work represent a special case of sparsely populated objects, we introduce a significant simplification to the imaging model by neglecting diffraction and refraction in order to preserve a reasonably complicated framework for computationally efficient image reconstruction.

Using the assumptions introduced above, the remaining radiation-matter interaction can be fully described by the complex and frequency dependent refractive index of the material
\begin{equation}
    \widetilde{n}\left(\omega\right)=n+i\kappa \,,
    \label{eq_refractiveIndex}
\end{equation}
where $n$ denotes the real refractive index and $\kappa$ denotes the absorption index. With the notation in Eq. (\ref{eq_refractiveIndex}), the measured THz signal, after travelling a distance $d$ through the object, can be written as
\begin{equation}
    E\left(d,t\right)=\int\limits _{-\infty}^{+\infty}\hat{E}^{\mathrm{ref}}\left(\omega\right)\exp\left[i\left(\omega t-\frac{\omega}{c}\left(\widetilde{n}\left(\omega\right)-n_{0}\right)d\right)\right]\mathrm{d}\omega \,.
    \label{eq_ModelSingleRay}
\end{equation}
Here $c$ refers to the speed of light in a vacuum, $n_0$ denotes the real refractive index of air and $\hat{E}^{\mathrm{ref}}$ is the Fourier transform of the reference signal $E^\mathrm{ref}$ measured in air. 

The complex refractive index $\widetilde{n}$ is commonly approximated by its Taylor series 

\begin{equation}
    \widetilde{n}\left(\omega\right) = \widetilde{n}\left(\omega_{0}\right)+\mathrm{\frac{d}{d\omega}}\widetilde{n}\left(\omega\right)|_{\omega_{0}} \left(\omega-\omega_{0}\right)+\mathcal{O}\left(\omega^{2}\right) \,.
    \label{eq_RefrIndexTaylor}
\end{equation}
A comparison with Fig. \ref{fig_SamplePreparation} shows that $n$ is only marginally varying over the frequency bandwidth of our measurement system and is therefore sufficiently approximated by a constant 0-th order term $\overline{n}=n\left(\omega_{0}\right)$.

We will further rewrite the absorption index $\kappa\left(\omega\right)$ by using the absorption coefficient \\ $\alpha\left(\omega\right)=2\kappa\left(\omega\right)\frac{\omega}{c}$. Fig. \ref{fig_SamplePreparation} shows that the frequency dependence of the absorption coefficient is low in contrast to other dielectric materials (e.g. ABS, PET \cite{Jin_Kim_Jeon_THzDielectricConstants_2006}). Low material parameter dispersion typically implies that the THz pulse shape is not significantly altered after propagating through the material. In Fig.~\ref{fig_ImagingSetup} a comparison of a reference pulse and a pulse after travelling through 2 mm of PP is shown, supporting the statement that the influence of dispersion is negligibly low. We therefore introduce the constant absorption coefficient $\overline{\alpha}=\alpha\left(\omega_{0}\right)$ as a further simplification to our model.

Inserting the average quantities $\overline{n}$ and $\overline{\alpha}$ into Eq. (\ref{eq_ModelSingleRay}) leads to

\begin{equation}
    E\left(d,t\right)=\int\limits _{-\infty}^{+\infty}\hat{E}^{\mathrm{ref}}\left(\omega\right)\exp\left[-i\frac{\omega}{c}\left(\overline{n}-n_{0}\right)d\right]\exp\left(-\frac{\overline{\alpha}}{2}d\right)\exp\left(i\omega t\right)\mathrm{d}\omega \,.
    \label{eq_FreqDomain_independentModel}
\end{equation}
The approximations introduced with $\overline{n}$ and $\overline{\alpha}$ allow us to transform our model back to the time-domain analytically. A more general approach would be to proceed with the frequency domain model in Eq. (\ref{eq_FreqDomain_independentModel}) in order to account for the frequency dependent material parameters. 

We leave this as subject to future work and continue by performing the inverse Fourier transform in equation \ref{eq_FreqDomain_independentModel}. This leads to the time-domain representation of the model

\begin{equation}
    E\left(d,t\right)=E^{\mathrm{ref}}\left(t-\frac{\overline{n}-n_{0}}{c}d\right)\exp\left(-\frac{\overline{\alpha}}{2}d\right) \,.
    \label{eq_timedomainModelSingleRay}
\end{equation}

Eq. (\ref{eq_timedomainModelSingleRay}) describes the interaction of a mathematical THz ray with a material of thickness $d$ and material parameters $\overline{n}$ and $\overline{\alpha}$. 
One can see that the interaction between sample and THz pulse is only expressed by a time-delay of the THz pulse with respect to the reference pulse, and an absorption term according to Lambert-Beers law.

In order to describe the CT measurement procedure (see section~\ref{subsec_CTMeasurements}), we introduce the discrete sample positions $s_i$ and projection angles $\theta_{j}$  for each point measurement, and the density function $f\left(x\right)$ representing the spatially dependent absorption coefficient. Analogously to our previous work \cite{Hubmer_Ploier_Ramlau_Fosodeder_vanFrank_2021}, we now introduce the Radon transform $R$ \cite{Natterer_2001} which is defined as the line integral along the line $L_{i,j}$
\begin{equation}
    \left(Rf\right)\left(s_{i},\theta_{j}\right)=\int_{L_{i,j}}f\left(x\right)\mathrm{d}x \,,
    \label{eq_RadonTransform}
\end{equation}
where the line $L_{i,j}$  is characterized by the sample position $s_i$ and the projection angle $\theta_{j}$.

In our specific use-case of extruded plastic profiles, the samples consist of only one material and air. The density function $f$ in Eq. (\ref{eq_RadonTransform}) is therefore piecewise constant with values either 0 or $\overline{\alpha}$. For a given $\left(s_{i},\theta_{j}\right)$, the Radon transform of $f$ is therefore proportional to the sample thickness $d_{i,j}$ along the line $L_{i,j}$
\begin{equation}
    \left(Rf\right)\left(s_{i},\theta_{j}\right)=\bar{\alpha}d_{i,j} \,.
    \label{eq_RadonTransform_binaryMaterial}
\end{equation}
We can use Eq. (\ref{eq_RadonTransform_binaryMaterial}) to rewrite Eq. (\ref{eq_timedomainModelSingleRay}) for the time domain THz-CT model
\begin{equation}
    E_{i,j}\left(t\right)=E^{\mathrm{ref}}\left(t-\frac{\bar{n}-n_{0}}{c\overline{\alpha}}\left(Rf\right)\left(s_{i},\theta_{j}\right)\right)\exp\left(-\frac{1}{2}\left(Rf\right)\left(s_{i},\theta_{j}\right)\right) \,,
    \label{eq_THzCTmodel}
\end{equation}
where $E_{i,j}$ denotes the final measurement signal in time-domain.

In the following, we deduce two different approaches for mapping each measurement signal $E_{i,j}(t)$ to a scalar sinogram value $g_{i,j}$ that acts as the basis for the image reconstruction approaches shown in section \ref{sec_ImageReconstructionoApproaches}.

\subsection{Pulse energy sinogram mapping}\label{subsec_PulseEnergySinogramMapping}

In the conventional approach of X-Ray CT, information about the sample is retrieved from the transmitted intensity of X-Ray radiation $I_{i,j}$ with respect to a reference intensity $I^{\mathrm{ref}}$. The sinogram values for X-Ray CT are given according to $g_{i,j}^X = - \ln\left[I_{i,j}/I^{\mathrm{ref}}\right]$.

Analogously, the energy of a THz pulse is defined as $\int_{-\infty}^{+\infty}E_{i,j}\left(t\right)^{2}\,\mathrm{d}t$. By calculating the energy from Eq. (\ref{eq_THzCTmodel}) we find

\begin{equation}
    g_{i,j}^E:=-\ln\left[\frac{\int\limits _{-\infty}^{+\infty}E_{i,j}\left(t\right)^{2}\,\mathrm{d}t}{\int\limits _{-\infty}^{+\infty}E^{\mathrm{ref}}\left(t\right)^{2}\,\mathrm{d}t}\right]=\left(Rf\right)\left(s_{i},\theta_{j}\right) \,,
    \label{eq_sinogramEnergy}
\end{equation}
with $g_{i,j}^E$ denoting the pulse energy sinogram values.

\subsection{Pulse phase sinogram mapping}\label{subsec_PulseDelaySinogramMapping}

The approach presented above extracts information about the sample based on the signal amplitude. Since the measurement setup is designed to measure the time evolution of the electric field of the THz pulse, it is crucial to find a way to access the phase information as well. According to our model in Eq. (\ref{eq_THzCTmodel}), the phase of the THz pulse is given by the time-shift term $\frac{\bar{n}'-n_{0}}{c\overline{\alpha}}\left(Rf\right)\left(s_{i},\theta_{j}\right)$ .

Standard signal processing offers a straightforward way to calculate the time-shift, for instance by means of peak-finder algorithms and thresholding. However, the resulting sinogram is non-continuous, which can lead to artifacts in the reconstructed image. Furthermore, this approach does not offer a robust way of dealing with THz signals containing multiple peaks. In practice, these signals typically appear due to the finite cross-section of the THz beam (FWHM: 1 mm) travelling through differently thick parts of the sample simultaneously. We have recently discussed this effect in \cite{Hubmer_Ploier_Ramlau_Fosodeder_vanFrank_2021} by means of a more complex, non-linear imaging model. 

In this work, we therefore proceed with our more compact model of the mathematical THz ray and extract the time-shift term in Eq. (\ref{eq_THzCTmodel})

\begin{equation}
\triangle t:=\frac{\bar{n}-n_{0}}{c\overline{\alpha}}\left(Rf\right)\left(s_{i},\theta_{j}\right) \,.
\label{eq_timeShiftTerm}
\end{equation}
Calculating the quantity $\int_{-\infty}^{+\infty}E_{i,j}\left(t\right)^{2}t\,\mathrm{d}t$ from Eq. (\ref{eq_THzCTmodel}) by substituting $t'=t-\triangle t$ results in 

\begin{equation}
\int\limits _{-\infty}^{+\infty}E_{i,j}\left(t\right)^{2}t\,\mathrm{d}t=\left[\int\limits _{-\infty}^{+\infty}E^{\mathrm{ref}}\left(t'\right)^{2}t'\,\mathrm{d}t'+\triangle t\int\limits _{-\infty}^{+\infty}E^{\mathrm{ref}}\left(t'\right)^{2}\,\mathrm{d}t'\right]\exp\left[\left(Rf\right)\left(s_{i},\theta_{j}\right)\right] \,.
\label{eq_substitutedIntegral}
\end{equation}
The remaining integrals in Eq. (\ref{eq_substitutedIntegral}) are independent of $f$ and calculated from the reference measurement. From Eq. (\ref{eq_sinogramEnergy}) we find that dividing Eq. (\ref{eq_substitutedIntegral}) by $\int_{-\infty}^{+\infty}E_{i,j}\left(t\right)^{2}\,\mathrm{d}t$ and rearranging terms results in 

\begin{equation}
\triangle t=\frac{\int\limits _{-\infty}^{+\infty}E_{i,j}\left(t\right)^{2}t\,\mathrm{d}t}{\int\limits _{-\infty}^{+\infty}E_{i,j}\left(t\right)^{2}\,\mathrm{d}t}-\frac{\int\limits _{-\infty}^{+\infty}E^{\mathrm{ref}}\left(t\right)^{2}t\,\mathrm{d}t}{\int\limits _{-\infty}^{+\infty}E^{\mathrm{ref}}\left(t\right)^{2}\,\mathrm{d}t} \,.
\label{eq_phaseInformation}
\end{equation}
Reinserting the definition Eq. (\ref{eq_timeShiftTerm}) back into Eq. (\ref{eq_phaseInformation}) and rearranging terms leads to the pulse phase sinogram values 

\begin{equation}
g_{i,j}^{P} :=\frac{c\overline{\alpha}}{\bar{n}-n_{0}}\left[\frac{\int\limits _{-\infty}^{+\infty}E_{i,j}\left(t\right)^{2}t\,\mathrm{d}t}{\int\limits _{-\infty}^{+\infty}E_{i,j}\left(t\right)^{2}\,\mathrm{d}t}-\frac{\int\limits _{-\infty}^{+\infty}E^{\mathrm{ref}}\left(t\right)^{2}t\,\mathrm{d}t}{\int\limits _{-\infty}^{+\infty}E^{\mathrm{ref}}\left(t\right)^{2}\,\mathrm{d}t}\right]=\left(Rf\right)\left(s_{i},\theta_{j}\right) \,.
\label{eq_sinogramPhase}
\end{equation}
The latter term in Eq. (\ref{eq_phaseInformation}) is calculated from the reference measurement and is therefore constant, while the material parameters cause a linear scaling of the sinogram values.

The advantage of this procedure over standard peak finding algorithms is its robustness and the absence of tuning parameters (e.g. thresholds). Furthermore, integrating over the THz signals has a time-domain averaging effect, leading to a continuous sinogram. In addition, it automatically takes into account the case of multiple peaks appearing in one THz signal, as discussed above.

\section{Image reconstruction approaches}\label{sec_ImageReconstructionoApproaches}

In this section, we shortly introduce the different image reconstruction approaches used throughout this work. Reconstructing based on both the pulse energy and the pulse phase mappings in Eq. \eqref{eq_sinogramEnergy} and \eqref{eq_sinogramPhase} amount to solving a linear system 
    \begin{equation}
        (Rf)(s_i,\theta_j) = g_{i,j} \,,
    \label{eq_Radon_discrete}
    \end{equation}
for the unknown density $f$, with right hand sides $g_{i,j} = g_{i,j}^E$ and $g_{i,j} = g_{i,j}^P$, respectively. Since Eq. \eqref{eq_Radon_discrete} is nothing else than a discretized version of the continuous Radon transform equation \cite{Natterer_2001,Louis_1989}
    \begin{equation}
        (Rf)(s,\theta) = g(s,\theta) \,,
        \label{eq_Radon_continuous}
    \end{equation}
standard reconstruction methods for this equation can be adapted to solve our THz-CT problem. Perhaps the most well-known of these methods is filtered back-projection, which is based on the classic Radon inversion formula \cite{Natterer_2001,Louis_1989}. The filtering step in this method acts as a regularization, which is necessary since the problem is in general ill-posed and unstable with respect to unavoidable measurement noise in the data. Another popular approach for solving Eq. \eqref{eq_Radon_continuous} is Tikhonov regularisation \cite{Engl_Hanke_Neubauer_1996}, which determines a stable approximation of $f$ by minimizing 
    \begin{equation}
        || Rf - g ||^2 + \beta ||f||^2 \,,
    \end{equation} 
where $\beta$ denotes a regularization parameter. The minimizer can be computed explicitly, in our discrete case as the solution of a linear matrix-vector equation. Alternatively, a common iterative approach for solving \eqref{eq_Radon_continuous} is given by Landweber iteration \cite{Engl_Hanke_Neubauer_1996}, which is defined by
    \begin{equation*}
         f_{k+1} = f_{k} + \gamma R^*(g - R f_k) \,,
    \end{equation*}
where $\gamma$ denotes a step-size parameter. Here, the choice of stopping index acts as regularization. 

For our THz-CT problem we focus on the above three reconstruction methods, i.e., on filtered back-projection, Tikhonov regularization, and Landweber iteration. For further details on these methods, as well as for other possible reconstruction methods we refer to \cite{Engl_Hanke_Neubauer_1996,Louis_1989}.

\section{Results}\label{sec_Results}

\begin{figure}[htbp]
    \centering
    \includegraphics[viewport=10bp 590bp 510bp 842bp,clip,width=10cm]{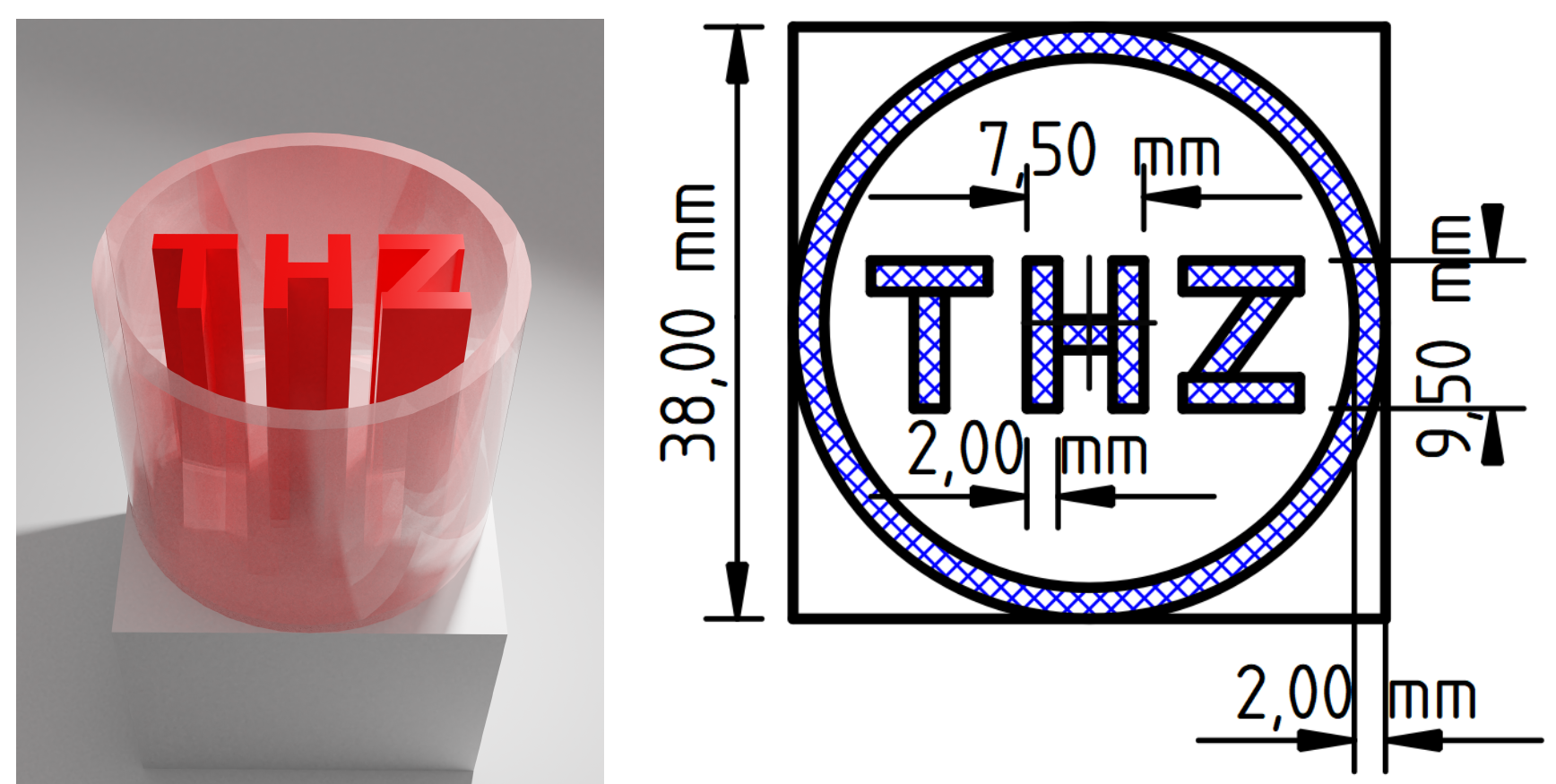}
    \caption{3D printed sample for THz CT measurements: 3D representation of a circular sample with the letters "THZ" inscribed (left). Cross-section of the sample (right). Cross-hatched areas represent solid walls of PP. }
    \label{fig_THZ_sample}
\end{figure}

In order to compare the outcomes of the two different sinogram mapping methods and reconstruction approaches presented above, we comparatively walk through and analyze the results obtained for the sample in Fig. \ref{fig_THZ_sample}. Further on, more reconstructed sample cross-sections, retrieved with the most promising combination of sinogram mapping and reconstruction approach, are shown below and in the additionally provided supplementary material (see \textcolor{urlblue}{Supplement 1}).

\subsection{Qualitative verification}\label{subsec_QualitativeVerification}

The purpose of the following measurements is to compare different methods of sinogram mapping. After choosing the most efficient method, different image reconstruction approaches are qualitatively evaluated. For this, the sample in Fig. \ref{fig_THZ_sample} was scanned over a range of 0 - 359\textdegree{} and the raw measurement data was preprocessed according to section~\ref{subsec_CTMeasurements}.

\begin{figure}[htbp]
    \centering
    \includegraphics[viewport=10bp 350bp 580bp 842bp,clip, width=12cm]{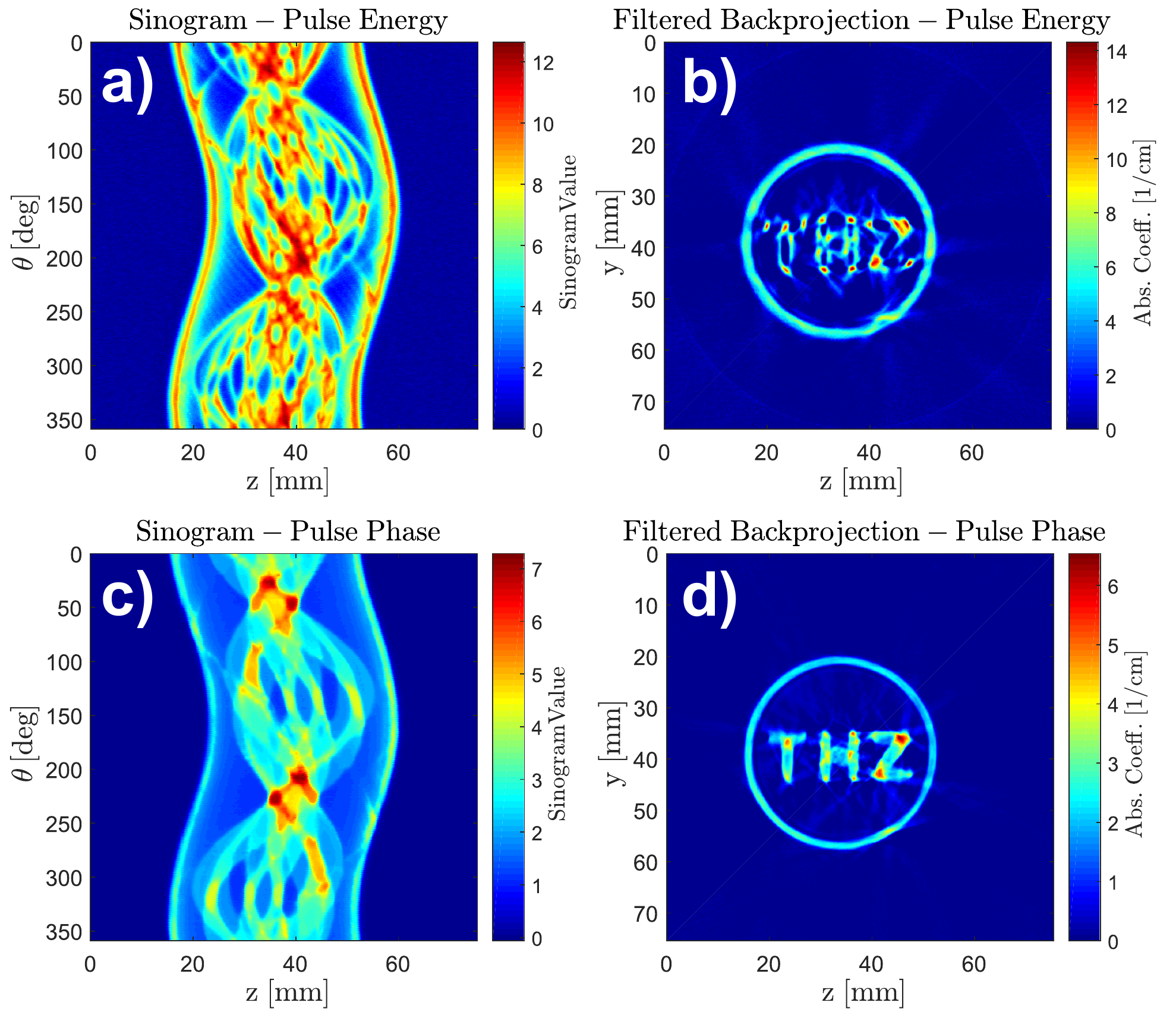}
    \caption{Comparison of sinogram mapping methods. (a) Pulse energy sinogram, (b) FBP of pulse energy sinogram, (c) Pulse phase sinogram, (d) FBP of pulse phase sinogram. The reconstructed image (d) obtained by pulse phase sinogram mapping (c) shows a significantly better representation of the sample geometry than conventional pulse energy based FBP (b).}
    \label{fig_CompareSinogramMapping}
\end{figure}

In section~\ref{subsec_PulseEnergySinogramMapping} and \ref{subsec_PulseDelaySinogramMapping}, two different approaches for extracting information from THz signals are discussed. For both approaches, the calculated sinograms are shown in Fig. \ref{fig_CompareSinogramMapping} (a, c). The reconstructed cross-sections in Fig. \ref{fig_CompareSinogramMapping} (b, d) were obtained by applying the classical filtered backprojection algorithm to the sinograms. Those image reconstructions were implemented using the built-in MATLAB \cite{MATLAB_2016} function \textit{iradon()} with default settings. The calculations were performed on a standard desktop PC (Intel i7-4790 CPU with 16 GB RAM) with a typical computation time of one second per reconstruction.

\begin{figure}[H]
    \centering
    \includegraphics[viewport=10bp 600bp 580bp 842bp,clip, width=12cm]{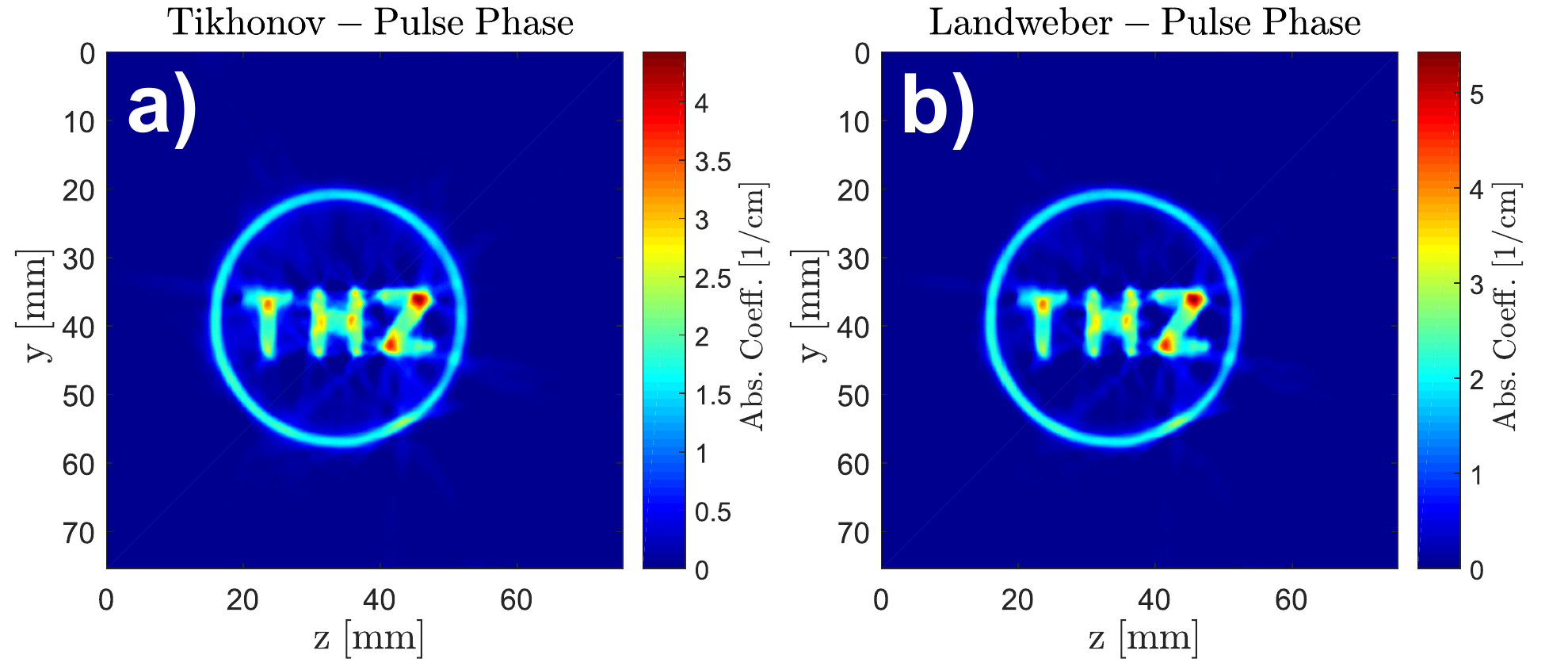}
    \caption{Comparison of image reconstruction approaches: (a) Tikhonov regularization, (b) Landweber iteration. Both reconstruction approaches yield similar results, with Landweber iteration (b) resulting in a more homogeneous representation of the letters \textit{THZ}.}
    \label{fig_CompareReconstructionApproaches}
\end{figure}

Using the pulse phase sinogram given in Fig.~\ref{fig_CompareSinogramMapping} (c), the different image reconstruction approaches discussed in section~\ref{sec_ImageReconstructionoApproaches} were tested in order to give a qualitative comparison between the filtered back-projection, Tikhonov regularization and Landweber iterations. The latter two were implemented in MATLAB using the AIR TOOLS II toolbox \cite{Hansen_Jorgensen_2017} for assembling the Radon transform matrix. Fig. \ref{fig_CompareReconstructionApproaches} shows the results of Tikhonov regularization (left) and Landweber iteration (right). These were obtained with the choice $\beta = 3700$ for the Tikhonov regularisation parameter and $\gamma = 3.8\cdot10^{-6}$ for the step-size in Landweber iteration, which was stopped after $70$ iterations. These parameters were optimized manually for minimal noise and maximum image sharpness. The total computation time was in the range of one minute on the same desktop PC as mentioned above. As no effort was made to increase computational efficiency, there is still potential to speed up the reconstruction algorithm.

\subsection{Quantitative evaluation of wall thicknesses}

For quantitative validation of our THz-CT measurements, we designed and measured two further samples with more ordinary geometries. Due to the sample design, scattering artifacts in the reconstruction are to be expected only due to 3D-printed inhomogenities (e.g. layer stacks, joints). 

\begin{figure}[H]
    \centering
    \includegraphics[viewport=50bp 345bp 580bp 842bp,clip, width=12cm]{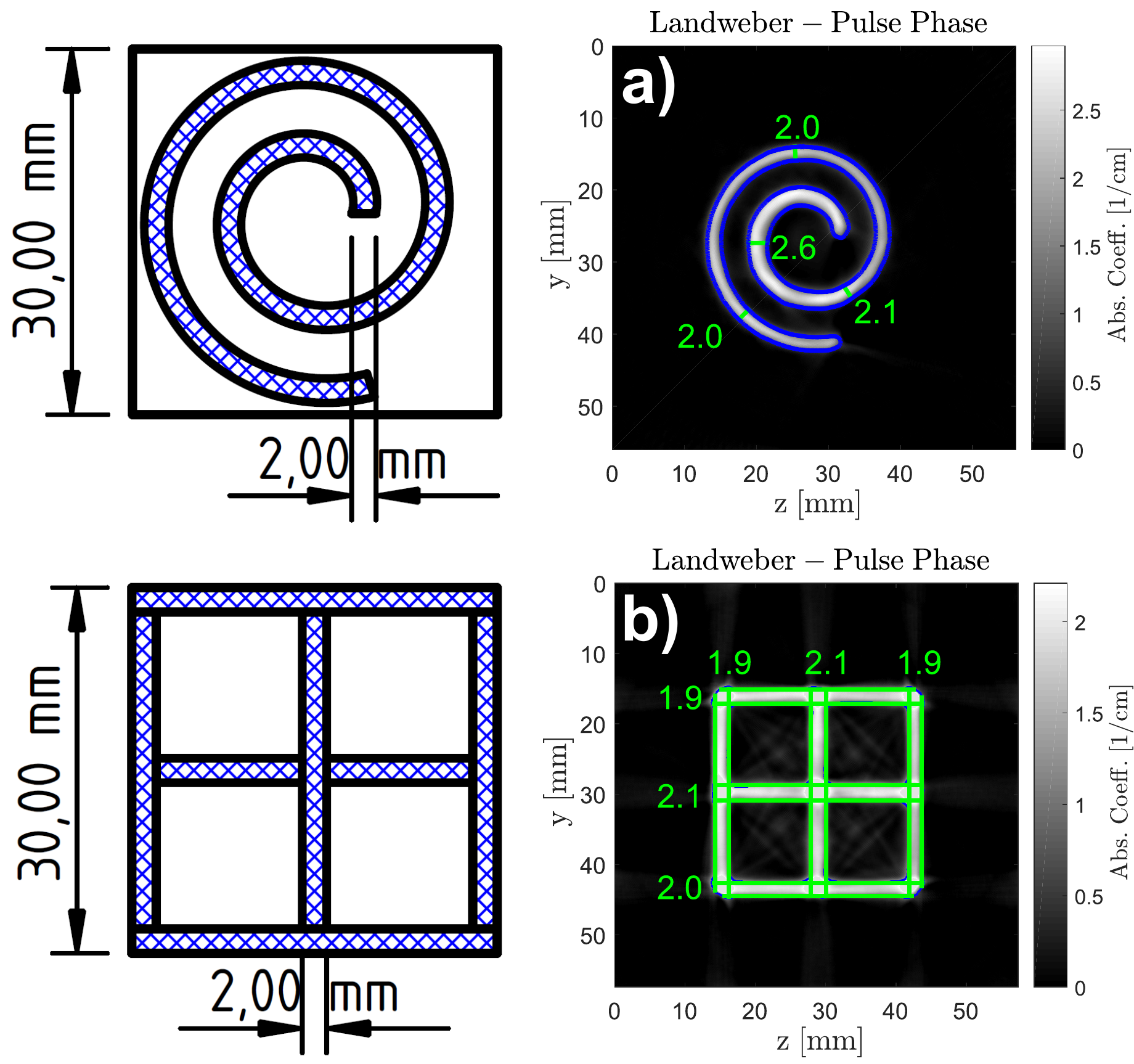}
    \caption{Quantitative evaluation of reconstructed images: Sample cross-sections (left), Quantitative evaluation of a curved sample (a), Quantitative evaluation of a planar sample (b). The blue hatched area in the sample cross-section represents the sample walls. Blue lines in the reconstructed image represent the calculated sample contour and green lines mark the identified sample boundaries using the Hough transform. Sample a) shows a gradually increasing deviation of the wall thickness towards its center of rotation. Overall good agreement within the accuracy of the 3d printer was achieved for sample b).}
    \label{fig_QuantitativeEval}
\end{figure}

While the geometry of the first object in Fig. \ref{fig_QuantitativeEval} (top) is parameterized by a spiral with linearly increasing radius, the second object (bottom) is highly symmetrical and consists of planar elements only.

Both images were reconstructed using Landweber iteration with the same parameters as in section~\ref{subsec_QualitativeVerification}. As a reference measurement for both objects, the nominal wall thickness (2 mm) was verified within minor deviations of $\pm0.1\,\mathrm{mm}$ by a mechanical caliper. The contours of the reconstructed objects were determined by calculating the iso-line (blue line) at the half maximum value of the reconstructed image. 

The wall thickness of the spiral was measured at discrete positions indicated by green lines. A variing thickness between 2.0 - 2.6 mm was found.  

Due to the linear geometry of the second object, we applied the linear Hough transform (LHT) to the contour of the object in order to identify a valid boundary along each sample wall. As a constraint for the LHT, only vertical and horizontal lines with a minimum length of 20 mm were accepted as valid object boundaries. The determined thicknesses  for the three horizontal and vertical wall segments are marked in Fig. \ref{fig_QuantitativeEval} (b).

\section{Discussion}

We begin this section with a qualitative comparison of the sinogram mapping methods in Fig. \ref{fig_CompareSinogramMapping}. In both cases (b, d), the cylinder around the letters \textit{THZ} is well identifiable. Since the cylinder is 3D printed from multiple circles next to and on top of each other, an inhomogenious joint/stack of layers is produced. This joint can be identified in both reconstructed images at the position $\left(y|z\right)\approx\left(55\,\mathrm{mm\,}|\,45\,\mathrm{mm}\right)$.

Generally, the pulse energy sinogram (a) and reconstruction (b) yields more pronounced edges and inhomogenities than the pulse phase sinogram. We presume that this is caused by the strong influence of scattering effects on small sample structures and edges, effectively reducing the detected signal amplitude. On the other hand, the pulse phase mapping approach eliminates the influence of the absolute signal amplitude by normalizing each measured signal with respect to its energy (see Eq. \ref{eq_sinogramPhase}). This property effectively suppresses the presence of large peaks in the reconstructed image (d), and leads to a better representation of the sample geometry in general. 

Furthermore, it is worth noting that the thickness of the letters in Fig. \ref{fig_CompareSinogramMapping} (d) is significantly larger than the cylinders wall thickness, despite both being printed with a thickness of $2.0\pm0.1\,\mathrm{mm}$ (measured with a mechanical caliper). We assume that this is a consequence of refraction in the cylinder, and point to further experimental evidence given below.

In general, for sparse sample structures with sharp edges we prefer the method of pulse phase mapping for geometry identification. For more specific applications, as for example detection of small inhomogenities in the dimension of the THz wavelength, pulse energy mapping might be superior.

In addition to the filtered backprojection image reconstruction method, we have qualitatively shown that Tikhonov regularization and Landweber iterations yield similar results (see Fig. \ref{fig_CompareReconstructionApproaches}). Especially within the letters \textit{THZ} we observe a much more homogeneous structure and less artifacts for Tikhonov regularization and Landweber iteration. After manual parameter optimization,  both reconstruction approaches seem to outperform the classical approach of filtered backprojection. 

We have further examined two more samples for quantitative testing of the accuracy of the proposed THz-CT approach. In Fig.~\ref{fig_QuantitativeEval} (a) we have observed a locally increasing wall thickness of up to 2.6 mm. The tendency indicates that the amount of surrounding sample structure layers artificially increases the wall thickness in the reconstructed image. Throughout this work, this effect was consistently and explicitly observed for all samples with curved surrounding structures (see \textcolor{urlblue}{Supplement 1}). Thus, we strongly assume that this artifact is predominantly caused by the wave propagation effect of refraction. According to geometrical optics, refraction on parallel, planar surfaces only applies a spatial shift to a ray, while refraction on a curved surface entirely changes the direction of the ray, thus introducing a larger inaccuracy of our forward model. So far, we have not accounted for this effect in our model. Nevertheless, the spiral structure is qualitatively well identifiable and we therefore plan to use the current approach as the basis for further improvements in the image reconstruction of curved objects. In the industrially more relevant case of a planar profile (see Fig.~\ref{fig_QuantitativeEval}, b), excellent agreement with the nominal wall thickness of $2.0\pm0.1\,\mathrm{mm}$ was achieved. The high accuracy of this result shows that our THz-CT approach is capable of detecting structural features down to industrially relevant dimensions.

Previously reported results obtained by THz-CT often involve samples with a very simple geometry (e.g. filled cylinders \cite{Mukherjee_Federici_pulsedTHzCT}). Measurements performed on slightly more complex samples \cite{Tripathi_THzCT_2016} often show unnatural structural artifacts (e.g. non-uniform wall thicknesses, unnatural edges), besides the effect of broadened walls, which is also reported in this work. Other reported systems \cite{Abraham_RefractionLossTHzCT} are very slow, resulting in long data acquisition times or poor image quality and thus are not useable in an industrial environment. To our knowledge, the THz-CT system presented here outperforms the systems reported in the literature in terms of data acquisition time, image quality and quantitative accuracy versus sample complexity, and robustness for different sample geometries.

\section{Conclusion}

In this work, we have presented two computationally efficient and robust approaches for performing THz-CT measurements on plastic profiles by means of a fast THz-TDS system. While the first approach, in the style of classical X-Ray CT, utilizes the amplitude information of the measured signals for image reconstruction, the second approach makes use of the availability of phase information in the THz-TDS measurement. We have experimentally shown that THz-CT based on the signal amplitude is prone to wave scattering artifacts. Structures with sizes in the dimension of the wavelength or sharp edges lead to extended peaks in the reconstructed image, potentially overshadowing other structures. In contrast, the influence of scattering is mostly eliminated in the second approach, leading to overall better representations of the sample geometry. 

We have further compared the performance of multiple image reconstruction methods, namely filtered backprojection, Tikhonov regularization, and Landweber iteration. It was empirically found that Tikhonov regularization and Landweber iteration yield more homogeneous images than filtered backprojection. 

As a last point, the quantitative performance of our THz-CT system was evaluated using the example of a curved and a planar sample (see Fig. \ref{fig_QuantitativeEval}). Local deviations were found for the special case of the curved sample. The planar sample could be reconstructed correctly within an accuracy of $\pm0.1\,\mathrm{mm}$. Due to the successful measurement on the planar sample and further collected experimental data (see \textcolor{urlblue}{Supplement 1}), we conclude that the artificially increased wall thickness observed for curved samples is a consequence of refraction. However, the quantitative accuracy of the reconstruction on the planar profile proves that our THz-CT approach is a valid tool for NDT on plastic profiles.

\begin{backmatter}
\bmsection{Funding}
Österreichische Forschungsförderungsgesellschaft (871974); European Commission (777222); Austrian Science Fund (F6805-N36).

\bmsection{Acknowledgments} 
Part of this work was performed under the scope of the COMET program within the research project "Photonic Sensing for Smarter Processes". This program is promoted by BMK, BMDW, the federal state of Upper Austria and the federal state of Styria, represented by SFG. This project was supported by the strategic economic- and research program "Innovative Upper Austria 2020" of the province of Upper Austria. "This work was supported by the Austrian Science Fund (FWF) in the project F6805-N36 within the Special Research Programme SFB F68: “Tomography Across the Scales”

\bmsection{Disclosures}
The authors declare no conflicts of interest.

\bmsection{Data availability} Data underlying the results presented in this paper are not publicly available at this time but may be obtained from the authors upon reasonable request.

\bmsection{Supplemental document}
See \textcolor{urlblue}{Supplement 1} for supporting content.
\end{backmatter}


\bibliography{mybib}






\end{document}